\documentclass[a4paper]{jpconf}

\usepackage{amsmath}
\usepackage{amssymb}
\usepackage{graphicx}
\usepackage{bm}
\usepackage{epsfig}
\usepackage{amsfonts}
\usepackage{color}
\usepackage{mathtools}

\usepackage{wrapfig}

\newcommand{\bsigma}{{\boldsymbol\sigma}}

\newcommand{\xv}{{\mathbf x}}

\newcommand{\bB}{{\bf B}}

\newcommand{\bDl}{{\bf\Delta}}
\newcommand{\bE}{{\bf E}}

\begin{document}

\title{In-medium heavy-quarkonium from lattice QCD spectral functions}

\author{Alexander Rothkopf}

\address{Institute for Theoretical Physics,  Heidelberg University, Philosophenweg 16, 69120 Heidelberg, Germany}

\ead{rothkopf@thphys.uni-heidelberg.de}

\begin{abstract}
We discuss recent progress in the study of in-medium heavy quarkonium using first-principles lattice QCD calculations and effective field theory. In particular our focus lies on real-time information carried by QCD spectral functions and we report on a novel strategy for their extraction from Euclidean-time simulations. Combining an effective description for heavy quarks with finite temperature lattice QCD we compute in-medium spectra providing insight in heavy quarkonium melting in a static thermal medium. On the other hand we use spectral information to extract the complex in-medium potential which provides the basis for a dynamical description of quarkonium real-time evolution via the Schr\"odinger equation.
\end{abstract}

\section{Introduction}

Heavy quarkonium, the bound state of a b or c-quark with its anti-partner, is a valuable probe of the strong interactions both in vacuum, as well as at high energy densities, such as in the droplet of quark-gluon-plasma created in relativistic heavy-ion collisions at RHIC and LHC \cite{Brambilla:2010cs}. 

Recent experimental studies of vector channel Bottomonium ($\Upsilon(nS)$) and Charmonium ($J/\Psi,\Psi(nS)$) by the CMS and ALICE collaborations have revealed two distinct faces of quarkonium, represented by the different flavors each. For $b\bar{b}$, a comparison of di-muon spectra in $p+p$ and $Pb+Pb$ collisions provided evidence for a pattern of sequential suppression, i.e. less deeply bound excited states are dissolved more easily than the ground state \cite{Chatrchyan:2011pe}. These findings are in line with the concept of Bottomonium forming early on in a collision and traversing the plasma as well defined probe. On the other hand ALICE observed \cite{Abelev:2012rv} at mid-rapidity that the suppression of Charmonium reduces with increasing centrality. This supports the picture of $J/\Psi$ formation from fully thermalized $c\bar{c}$ at the phase boundary, as advocated for in the statistical model of hadronization \cite{Andronic:2009sv}. Understanding the intricate in-medium physics leading to each of the two scenarios from first principles has to be the central goal for theory.

Heavy quarkonium is characterized by the well pronounced separation of scales between the constituent rest mass ($m_c=1.275(25)$GeV, $m_b=4.66(3)$GeV) \cite{Agashe:2014kda} and all other scales present in current experiments. This distinguishes them as well defined observable, measurable with high precision, and opens up the possibility for theory to employ effective descriptions that allow a more intuitive understanding of the bound state physics in the presence of a thermal medium.

Already at zero temperature it is the intrinsic scale of quantum fluctuations of QCD, often referred to as $\Lambda_{\rm QCD}\simeq 200$MeV that lies well below $m_Q$, which in turn tells us that quantum fluctuations cannot spontaneously create a $Q\bar{Q}$ pair. Such a system, where creation and annihilation do not play a central role can be described by a non-relativistic effective field-theory (EFT) called NRQCD, in fact one can show that it can even be described to a good degree by a non-relativistic potential entering a Schr\"odinger equation \cite{Brambilla:2004jw}. At finite temperature the additional scale $T$ enters and one might ask how its presence changes the properties of the bound state, which eventually will melt if thermal fluctuations overcome its typical binding energy. NRQCD when used at T=0 only requires a separation of $m_Q$ from the next lower scale and thus in principle is valid also in-medium. We report on its application to lattice QCD studies of heavy quarkonium at $T>0$ in more detail in sec.\ref{sec3}.

On the other hand, the question whether an in-medium potential $V(r)$ can be defined and what its values are, has only recently been answered in a satisfactory way. For more than two decades, in the absence of the means to derive a potential from QCD, theorists have resorted to model potentials, among which two candidates have gained popularity. The purely real color singlet free energies in Coulomb gauge $e^{F^{(1)}(\mathbf{x})}=\langle {\rm Tr}\Big[\Omega(\mathbf{x}) \Omega^\dagger(0)\Big]\rangle_{\rm CG}$ from Polyakov-loop $\Omega(\mathbf{x})={\rm exp}\Big[ig\int_0^{\beta} d\tau A^4(\tau,\mathbf{x})\Big]$ correlators \cite{Nadkarni:1986as} and the internal energies $U^{(1)}=F^{(1)}-TS$ derived from it  \cite{Satz:2008zc}, as well as linear combinations \cite{Wong:2004zr} have been employed.

It is important to note that while these quantities are readily accessible in lattice QCD \cite{Kaczmarek:2007pb} they have not been linked to a potential in QCD. In fact progress in the formulation of EFT's \cite{Barchielli:1986zs,Brambilla:2008cx} made it possible to actually derive an expression for the potential between two static quarks in a thermal medium starting from the QCD Lagrangian, revealing that neither $F^{(1)}$ nor $U^{(1)}$ alone can play the role of $V(r)$. 

The reason is that the evaluation of the potential both in resummed hard-thermal loop perturbation theory \cite{Laine:2007qy,Beraudo:2007ky} and on the lattice \cite{Burnier:2014ssa,Burnier:2014yda} showed  that $V(r)$ is actually a complex quantity. While at $T>T_C$, ${\rm Re}[V]$ is closely related to the phenomenon of Debye screening, ${\rm Im}[V]$ is intimately linked to the scattering of the medium partons with the color string spanning in between the $Q\bar{Q}$ \cite{Beraudo:2007ky, Brambilla:2008cx}. Using a purely real model potential neglects these effects and necessarily misses an important aspect of the in-medium physics of heavy quarkonium. Recent lattice QCD results on the extraction of the complex in-medium potential are presented in sec.\ref{sec4}.

While the non-relativistic character of the heavy quarks allows a simplification of their description, the confinement aspects of the QCD vacuum as, well as the physics of the strongly coupled QGP at $T\simeq T_C$ require fully relativistic, as well as non-perturbative methods. One truly first principles approach is lattice QCD, where spacetime is discretized on a hyper-cubic grid with finite lattice spacing $a$ and box size $N_x a$. Gauge fields denoted in continuum by $A_\mu^a$ are represented by SU(3) valued link variables $U_\mu(x)={\rm exp}\big[igaA_\mu^a T^a\big]$, where $T^a$ represent the generators of the gauge group. These object reside on the links connecting the nodes of the spacetime grid. The light fermions of the QCD medium on the other hand populate the nodes. 

The strength of the lattice approach lies in the fact that the Feynman path integral of this theory can be evaluated non-perturbatively via stochastic Mont-Carlo methods, after rotating the Minkowski-time axis onto imaginary time. Since the simulation is not carried out in physical time a form of analytic continuation is required if dynamical information is to be extracted. This represents an inherently ill-defined problem. One strategy, Bayesian inference, proven successful in many areas of physics to overcome such a challenge is discussed in sec.\ref{sec2}.

At $T=0$ the combination of lattice QCD and NRQCD, have been highly successful in reproducing the quarkonium spectrum (see e.g. \cite{Thacker:1990bm,Dowdall:2011wh}). There is no modeling input necessary in these computations, thus they truly link the QCD Lagrangian to physically observable quantities, such as particle masses or matrix elements. Note that since all simulations are carried out using dimensionless numbers, at some point experimental input is required to set the overall scale. Once the scale is fixed however ab-initio predictions are possible, one example being the existence of the $\eta_b(2S)$ state with mass $m_{\eta_b(2S)}^{\rm LNRQCD}=9.988(3)$GeV predicted in 2011 \cite{Dowdall:2011wh}, which was confirmed by the Belle collaboration \cite{Mizuk:2012pb} at $m_{\eta_b(2S)}^{\rm Belle}=9.999(3.5)(2.8)$  in the following year.

At finite temperature lattice QCD can also help us to elucidate the static properties of the QCD medium, now that consistent continuum extrapolated results for thermodynamic quantities, such as the energy density and pressure have been obtained \cite{Bazavov:2011nk,Borsanyi:2013bia,Bazavov:2014pvz}. While the transition between the confined and deconfined phase is a mere crossover at physical quark masses \cite{Aoki:2006we}, a pseudo-critical temperature of $T_C=155(9)$MeV has been established that characterizes the change from the hadronic to QGP regime. Evaluation of the trace anomaly \cite{Borsanyi:2013bia,Bazavov:2014pvz,Burger:2014xga}, a non-perturbative measure of the interaction strength revealed a region of strong coupling  exists in the regime just above $T_C$. This reinforces the need for non-perturbative methods when approaching the physics of heavy quarkonium in heavy-ion collisions.

\section{Spectral functions from lattice QCD}
\label{sec2}

Physical information relevant for an understanding of in-medium heavy quarkonium can be found in spectral functions $\rho(\omega)$. They inform us about the presence and strength of bound states, which appear as peaked features of skewed Breit-Wigner type. The position encodes the mass of the state, the width its transition probability to other bound states or into the continuum, which is manifested as a featureless structure at large $\omega$. What makes spectral functions interesting from a technical point of view is that they constitute a bridge between correlation functions in real-time and those analytically continued to Euclidean time, which will play a crucial role in the extraction of the heavy quark potential from lattice QCD simulations in sec.\ref{sec4}.

In sec.\ref{sec3} we will calculate heavy quarkonium correlation functions $D(\tau)$ on the lattice. These are related to spectra via a Kernel $K(\tau,\omega)$, which is determined by QCD and exactly known\vspace{-0.3cm}
\begin{align}
 D(\tau,\mathbf{x})=\int\,d\omega\, K(\tau,\omega)\, \rho(\omega,\mathbf{x}),\quad D(\tau_i)=\sum_{l=1}^{N_\omega} \,\Delta\omega_l \, K_{il}\, \rho_l.\label{Eq:PropSpec}
\end{align}
In both studies presented in sec.\ref{sec3} and sec.\ref{sec4} we will encounter the $T$ independent Kernel $K(\tau,\omega)=e^{-\omega\tau}$, with which the above expressions amount to Laplace transforms.

Writing the convolution on the right of Eq.\eqref{Eq:PropSpec} with $N_\tau$ steps along imaginary time and $N_\omega$ frequency points, reveals the challenge we face in inverting this relation. I.e. a function $\rho$ with an intricate set of features has to be computed from a finite set of noisy datapoints obtained from a Monte Carlo simulation. This is an ill-defined problem, since for any discretization $\tau_i=\beta/N_\tau$ and in the presence of inevitable noise on $D$, an infinite number of choices for the $\rho_l$'s exist that will reproduce the data, even in the presence of the unitarity constraint $\rho_l>0$. 

One possibility to give meaning to the problem is through so called Bayesian inference, which is based on Bayes theorem 
 $P[\rho|D,I]\propto P[D|\rho,I]P[\rho|I]$.
It tells us that the probability for a test function $\rho$ to be the correct spectral function, given the simulated lattice data $D$ and any further prior information $I$ on the spectrum, consists of essentially two terms. The likelihood probability $P[D|\rho,I]=e^{-L}$ denotes the probability of the lattice data if $\rho$ were the correct spectrum and $L$ amounts to nothing but the usual $\chi^2$ quadratic distance functional. The second term $P[\rho|I]=e^{S}$ denotes the prior probability, which encodes prior information on the spectrum and acts as regulator of the degenerate $\chi^2$ fit. On the one hand the functional form of $S$ encodes part of $I$, on the other hand $S[\rho,m]$ depends on a function $m(\omega)$, the default model, which defines the extremum of $S$ and thus represents the correct spectrum in the absence of data. We have to carry out a numerical optimization to find the most probable $\rho$ given $D$ and $I$, with the competing $L$ and $S$ leading to a unique answer. 

One popular implementation of the Bayesian strategy is the Maximum Entropy Method (MEM) \cite{Asakawa:2000tr,Jakovac:2006sf,Rothkopf:2011ef}. In it the Shannon-Jaynes entropy is used as regulator term $S=S_{\rm SJ}$. One difficulty lies in the fact that $S_{\rm SJ}$ contains flat directions, which can adversely affect the convergence properties of the method and in practice necessitates the use of a priori restrictions of the functional space from which the spectra can be chosen. To avoid this resolution impediment we instead use a recently developed Bayesian approach \cite{Burnier:2013nla} that relies on a prior $S$ derived from the following three physical consideration: positivity; smoothness of the spectrum, where data has not introduced sharp features and the independence of the result from the units chosen. As discussed in \cite{Burnier:2013nla} these lead us to a prior functional 
$ S=\alpha \sum_{l=1}^{N_\omega}\,\Delta\omega_l\Big( 1-\frac{\rho_l}{m_l}+{\rm log}\Big[\frac{\rho_l}{m_l}\Big]\Big)$,
which is devoid of the flat directions of $S_{\rm SJ}$. Note that the weighting factor $\alpha$ here in contrast to the MEM can actually be integrated out explicitly in a Bayesian fashion \cite{Burnier:2013nla}. With $S$ set, we only need to specify $m(\omega)$, for which we will use a constant, the most neutral choice.


\section{Bottomonium spectra from lattice NRQCD}
\label{sec3}

Let me paint in broad strokes the ingredients to investigating heavy-quarkonium in the non-relativistic field theory NRQCD on the lattice at finite temperature using Bottomonium as example. Readers interested in more technical details can consult e.g. \cite{Thacker:1990bm}. Due to the separation of scales between heavy-quark rest mass and the two other characteristic scales $\Lambda_{\rm QCD}$ and $T$, creation and annihilation of heavy quark pairs is highly suppressed. This allows us to go over from a description of the quark fields in terms of four-component Dirac spinors $Q(x)$ to a set of two two-component Pauli spinors $(\psi(t,\mathbf{x}),\chi(t,\mathbf{x}))$ representing the quark and anti-quark contribution respectively. Formally this separation of the upper and lower components of the four-spinor is obtained from a systematic expansion of the QCD Lagrangian in terms of increasing powers of the inverse rest mass (or more precisely the heavy quark velocity $v$) via the Foldy-Tani-Wouthuysen transformation. This is the same approach used to derive the Hydrogen energy levels from the Dirac equation.

If we place the non-relativistic theory on a Euclidean spacetime lattice, the Pauli spinors obey the corresponding non-relativistic Hamiltonian, \vspace{-0.25cm}
\begin{align} 
H= H_0+ \delta H = - \frac{\Delta^{(2)}}{2M_b} - \frac{g}{2 M_b} \bsigma\cdot\bB  \hm + \frac{ig}{8 M_b^2}
(\bDl^{\pm}\cdot \bE - \bE\cdot \bDl^{\pm}) + \ldots
\end{align}
Here and $\Delta^\pm$ and $\Delta^{(2)}$ denote the lattice derivative and Laplace operator respectively. If all factors of the lattice spacing $a$ were made explicit it turns out that the expansion parameter amounts to $aM_b$ on the lattice. For the UV behavior to be well defined its value must be larger than $1.5$ \cite{Davies:1991py}, which reflects the fact that NRQCD works better on coarser lattices. 

The fact that quark and antiquark modes are separately propagating in the medium background leads to an initial-value problem for the heavy quark propagator $G (\xv, \tau_i )=\langle\psi(\xv, \tau_i )\psi(0,0)\rangle$ in the background of medium fields. Written in terms of the link variables provided as external fields by conventional lattice simulations we have
\begin{align}
G (\xv, \tau_i ) = \hm  \left(1 - \frac{H_0}{2n}\right)^n U_4^\dagger(\xv, \tau) \left(1 - \frac{H_0}{2n}\right)^n \nonumber \times \left(1 -\delta H \right)  G (\xv, \tau_{i-1})
\end{align}
The parameter $n$ reflects a freedom in the choice of the time discretization and is set to $n=2$ here. From $G (\xv, \tau_i )$ we can construct heavy quarkonium propagators 
$D({\mathbf x},\tau)=\sum_{{\mathbf x}_0}\langle O({\mathbf x},\tau)
G({\mathbf x},\tau) O^\dagger({\mathbf x}_0,\tau_0) G^\dagger({\mathbf
  x},\tau) \rangle_{\rm med}$   
selecting the individual channels by appropriate vertex operators $O(^3S_1;{\mathbf x},\tau)=\sigma_i$ for the S-wave $O(^3P_1;{\mathbf x},\tau)=\overset{\leftrightarrow_s}{\Delta}_i\sigma_j-\overset{\leftrightarrow_s}{\Delta}_j\sigma_i$ for the P-wave.

\begin{figure}
\centering
 \includegraphics[scale=0.39]{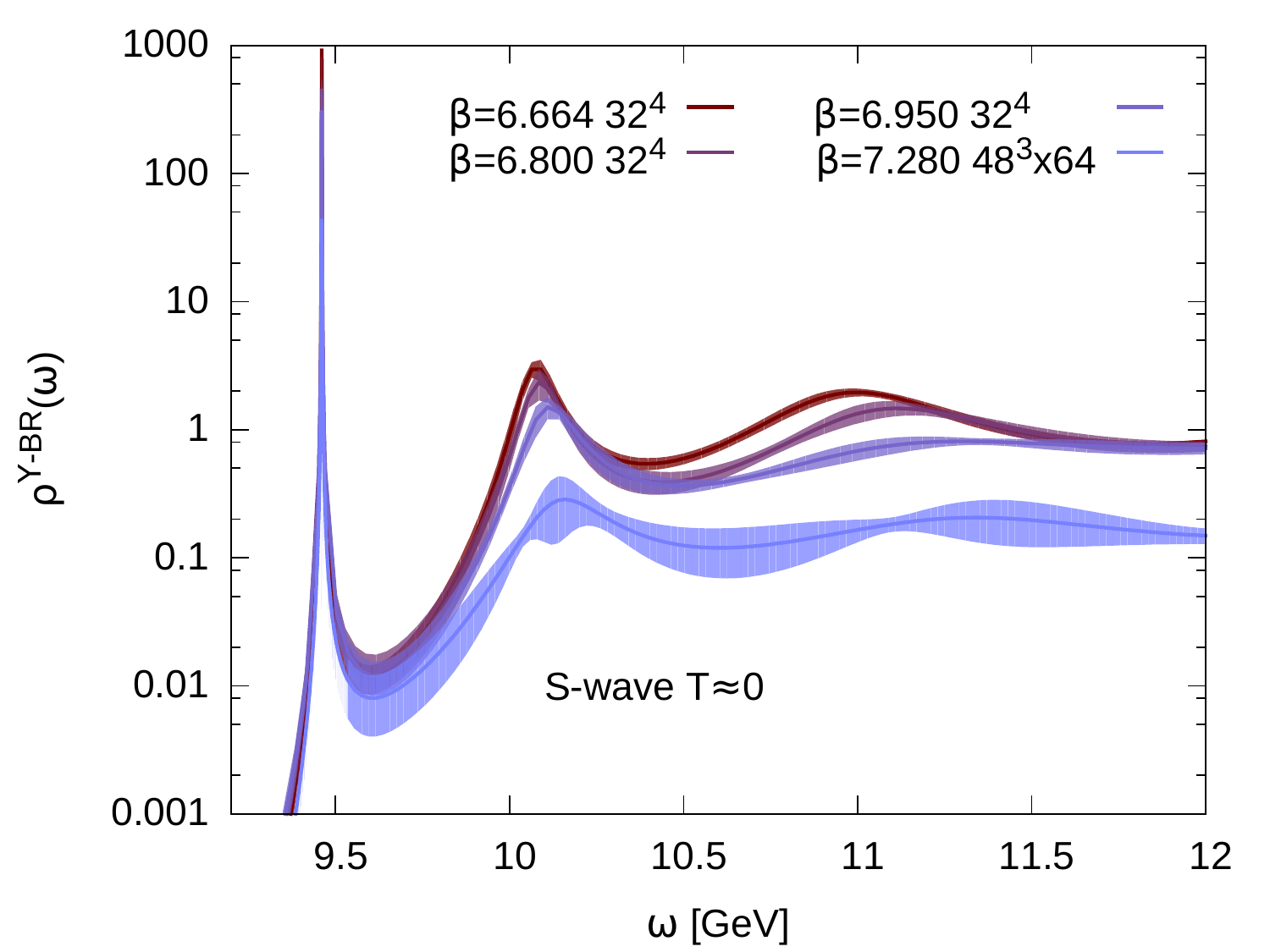}\hspace{1cm}
 \includegraphics[scale=0.39]{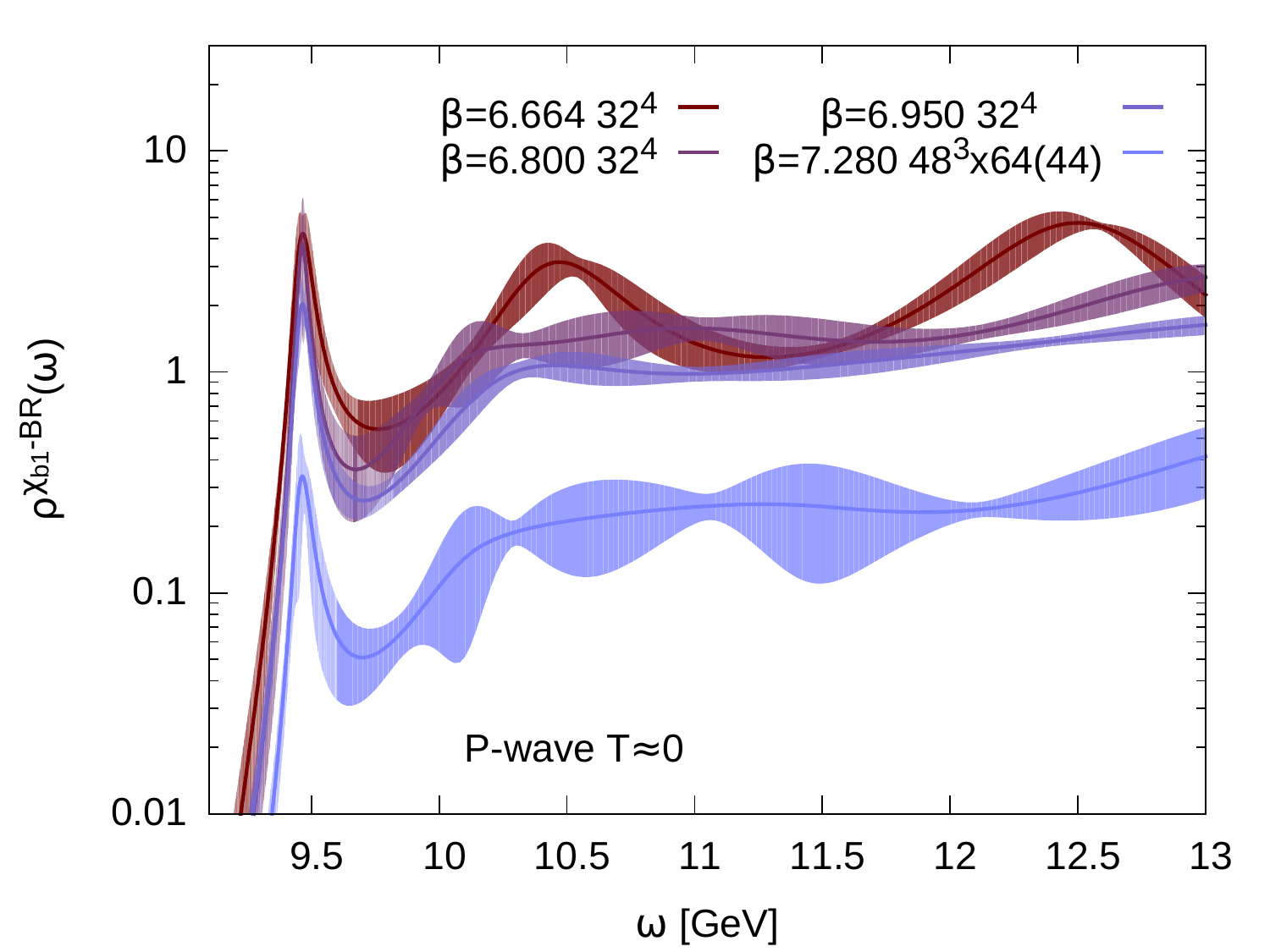}\vspace{-0.32cm}
 \caption{S-wave (left) and P-wave (right) Bottomonium spectra at $T\approx0$.}\label{Fig:LowTBottom}
\end{figure}

\begin{figure}\vspace{-0.32cm}
\centering
 \includegraphics[scale=0.45]{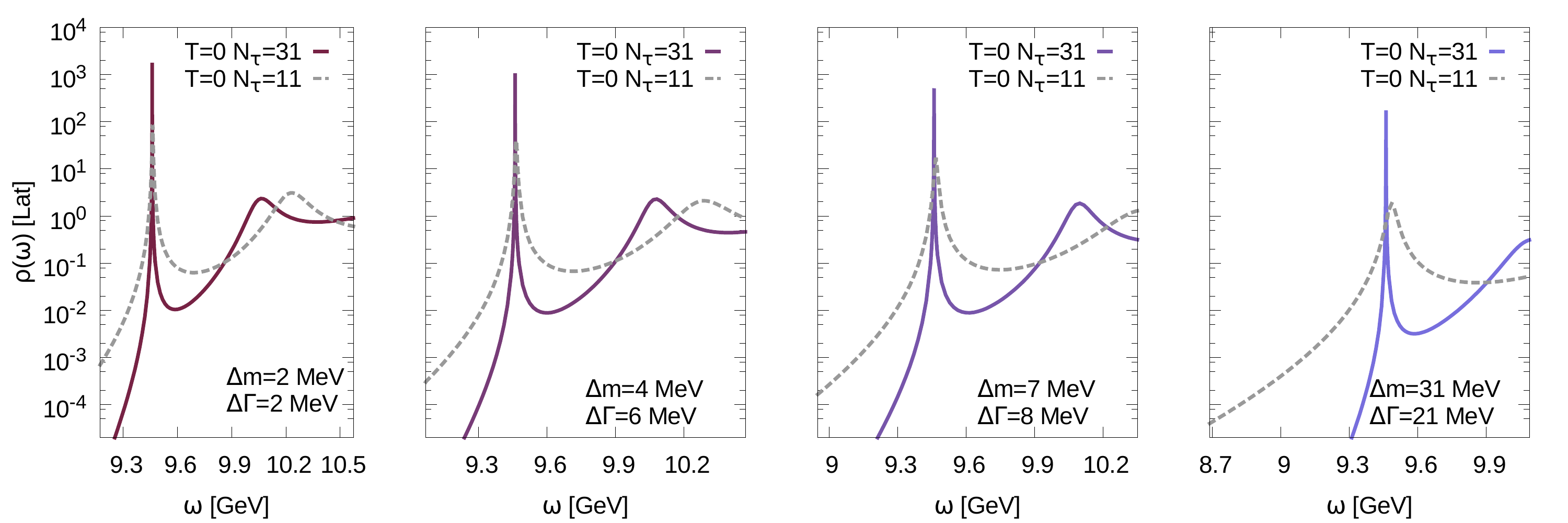}\vspace{-0.32cm}
 \caption{Spectra of the full (solid) and truncated (dashed) $T=0$ correlator sets.}\label{Fig:TestingRec}
\end{figure}

The medium expectation value is obtained by measuring and averaging these correlators on realistic gauge field configurations containing light u,d, and s (HISQ) quarks, generated by the HotQCD collaboration \cite{Bazavov:2011nk}. These $48^3\times 12$ lattices with $T_C=159MeV$ feature an almost physical pion mass of $m_\pi=161$MeV and allow a fine temperature scan between $T=140-249$MeV, since the lattice spacing is used to regulate the extend in Euclidean time  $\beta=6.664-7.280$. The expansion parameter lies between $aM_b=2.76-1.56>1.5$ allowing for the application of NRQCD. For a series of studies using a complementary fixed scale approach see \cite{Aarts:2011sm}. Experimental input, e.g. the $\Upsilon(1S)$ mass is required to set the relative and absolute frequency scale, for which we use low temperature ensembles with $48^3\times48,64$ .

At $T\approx0$ we have extracted \cite{Kim:2014iga} (left) the S- and (right) P-wave spectra shown in Fig.\ref{Fig:LowTBottom}. The S-wave ground state peak is very well resolved, around an order of magnitude more clearly than what is possible with the MEM. The second bump contains mostly contributions from $\Upsilon(2S)$, the higher lying wiggles are numerical ringing. After fixing the scale by the S-wave, we can estimate the systematic error in the computation from our postdiction for the $\chi_b(1S)$ mass. We obtain a value of $m_{\chi_b(1P)}^{\rm NRQCD}=9.917(3)$GeV slightly larger than the PDG value of $m_{\chi_b(1P)}^{\rm PDG}=9.89278(26)(31)$GeV. The P-wave ground state appears more washed out than the S-wave, not only because  of a lower signal to noise ratio in the underlying correlators. Even in continuum the ratio of P-wave peak height to background is smaller in comparison \cite{Burnier:2007qm}.

To investigate in-medium modification we have to ascertain how well our spectral reconstruction fares at $T>0$. The difference to vacuum lattices $N_\tau=48$ is the physical extend in Euclidean time, reflected in a smaller number of $N_\tau=12$ correlator points at $T>0$. Hence we take the $T=0$ correlator data set and truncate it after twelve points ($K(\omega,\tau)$ is $T$ independent). Supplying it to the reconstruction yields the spectra shown as gray dashed curves in Fig.\ref{Fig:TestingRec}. On the coarsest lattices ($\beta=6.664$) where NRQCD works best, the ground state peak reconstruction remains unaffected ($\Delta m<2{\rm MeV}$,$\Delta \Gamma<2{\rm MeV}$), while on the finest lattices ($\beta=7.280$) corresponding to the highest temperature in the following, we find ($\Delta m<40{\rm MeV}$,$\Delta \Gamma<21{\rm MeV}$). As expected for the P-wave the accuracy limits are even larger \cite{Kim:2014iga}. Note that the changes are on the level of individual MeV, which could not have been resolved in the same manner with the MEM. If at finite temperature we were to find changes larger than the stringent limits set here, we can attribute them to in-medium modification.

\begin{figure}
\centering
 \includegraphics[scale=0.27,angle=-90]{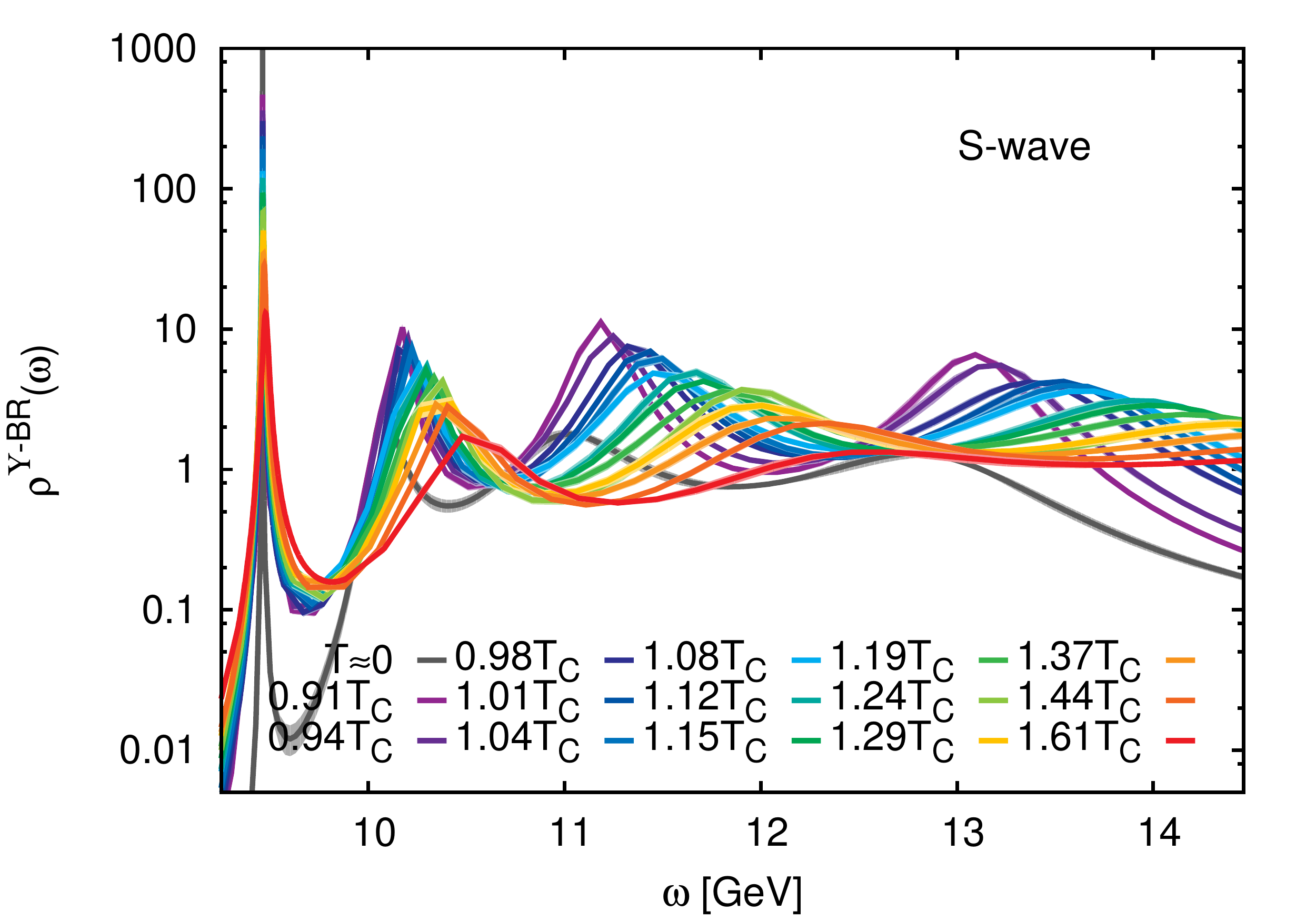}\hspace{1cm}
 \includegraphics[scale=0.27,angle=-90]{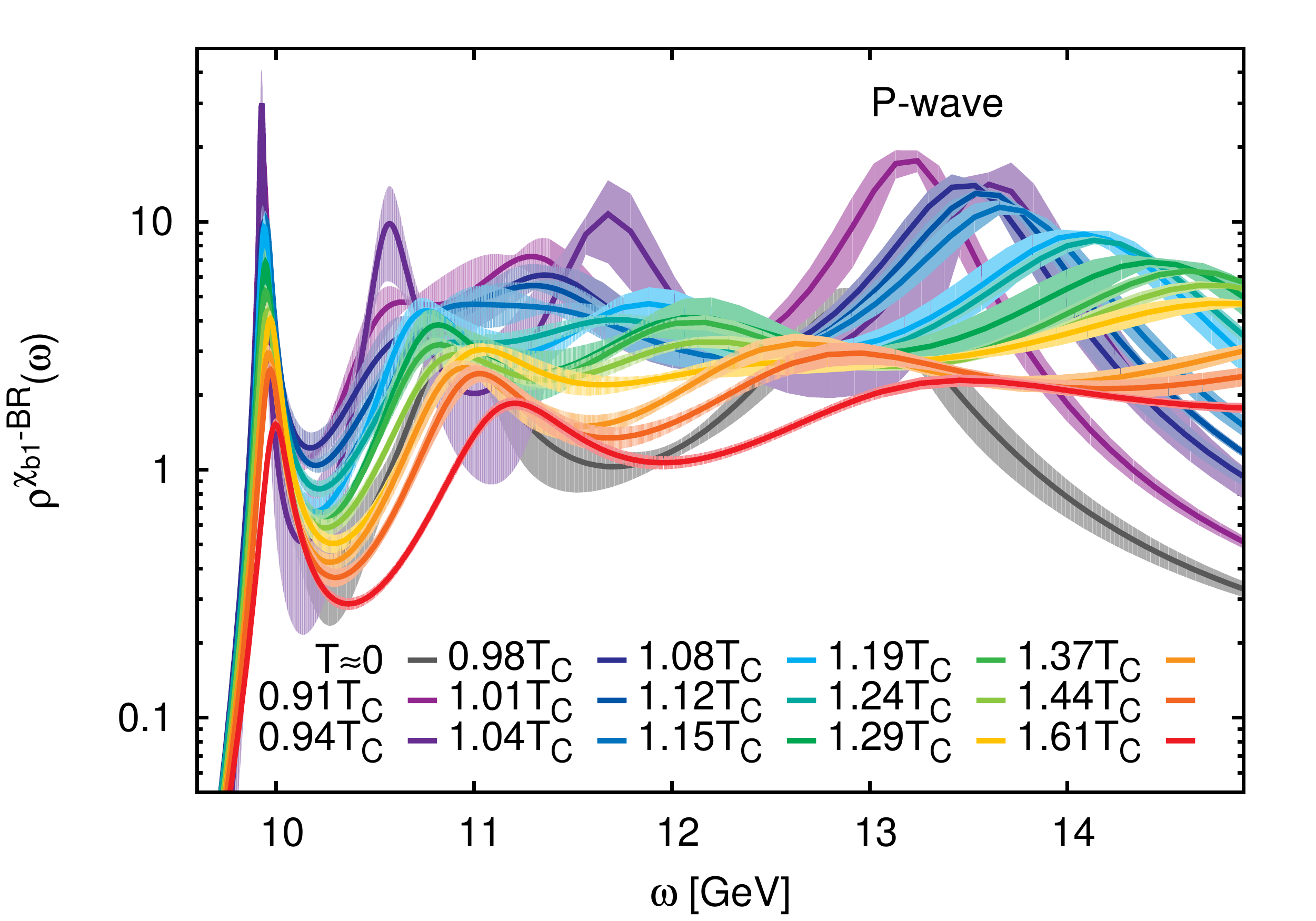}\vspace{-0.2cm}
 \caption{S-wave (left) and P-wave (right) Bottomonium spectra at $T>0$.}\label{Fig:FiniteTSpec}\vspace{-0.4cm}
\end{figure}

In Fig.\ref{Fig:FiniteTSpec} we show the $T>0$ results for (left) the S-wave and (right) the P-wave spectra at the fourteen different temperatures available. In the S-wave we find at all $T$ a well defined ground state peak, much larger than any further numerical ringing at higher $\omega$. If zoomed in, the peak is seen to slightly shift and broaden with $T$, however the changes are smaller than the accuracy limits established before. Hence we refrain from attributing them to medium effects. Nevertheless by a simple inspection by eye it is evident that $\Upsilon(1S)$ survives up to $T=249$MeV. This conclusion agrees with the results from recent a MEM based study \cite{Aarts:2011sm}.

The P-wave case is more subtle, not only because of the larger errors due to the lower signal to noise ratio in the correlators. While we find a well defined lowest lying peak at all $T$ its height is comparable to that of higher lying numerical ringing at the highest $T$. A naive inspection by eye fails to give conclusive insight. Therefore we propose to put the decision of survival or melting on more systematic footing via a comparison with non-interacting spectral functions. 

\begin{wrapfigure}{r}{0.5\textwidth}
  \begin{center}\vspace{-0.6cm}
   \includegraphics[scale=0.32]{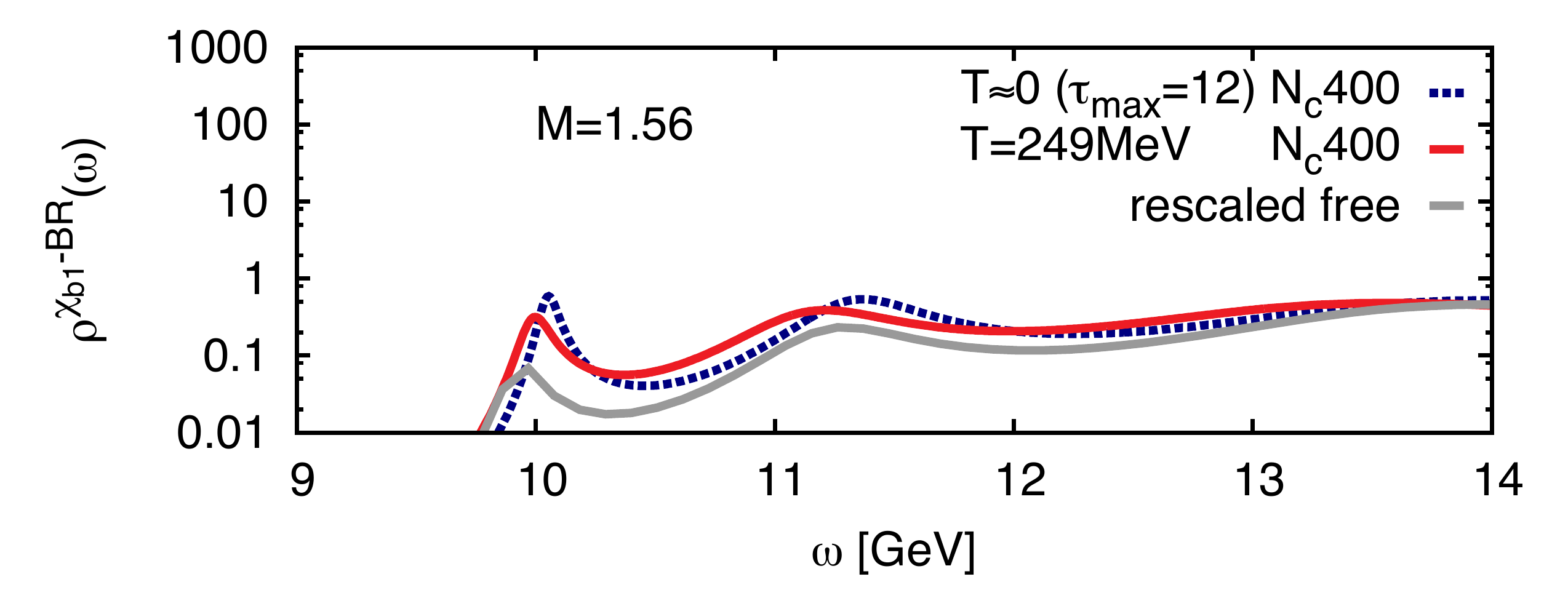}\vspace{-0.8cm}
  \end{center}
 \caption{Comparison of free (gray), interacting (red) and the spectral function from the truncated $T=0$ dataset (blue dashed)}\label{Fig:CompFree}\vspace{-0.4cm}
\end{wrapfigure}The $Q\bar{Q}$ spectrum in the absence of interactions is analytically known, not to show any bound state peaks. Such spectra can also be extracted numerically from lattices with all links set to unity and subsequent Bayesian reconstruction. Just as with Gibbs ringing in the Fourier transform we expect wiggly structures to appear, even if absent in the actual free spectrum. This can be seen in the gray curve in Fig.\ref{Fig:CompFree} where the free spectrum has been shifted to match the $\omega$ range of the full spectrum (red). We see that while numerical ringing dominates the second bump, the first peak is a factor of three larger and thus constitutes a signal for bound state survival. Hence we conclude that $\chi_b(1P)$ survives up to $T=249$MeV ($1.61T_C$). This conclusion differs from a recent MEM based study, which found melting already at $T=1.27T_C$ \cite{Aarts:2011sm}. One explanation might lie in the restricted search space of the MEM, which shrinks with the number of available datapoints and thus entangles physical effects and numerical limitations.

\begin{figure}
\centering
 \includegraphics[scale=0.4]{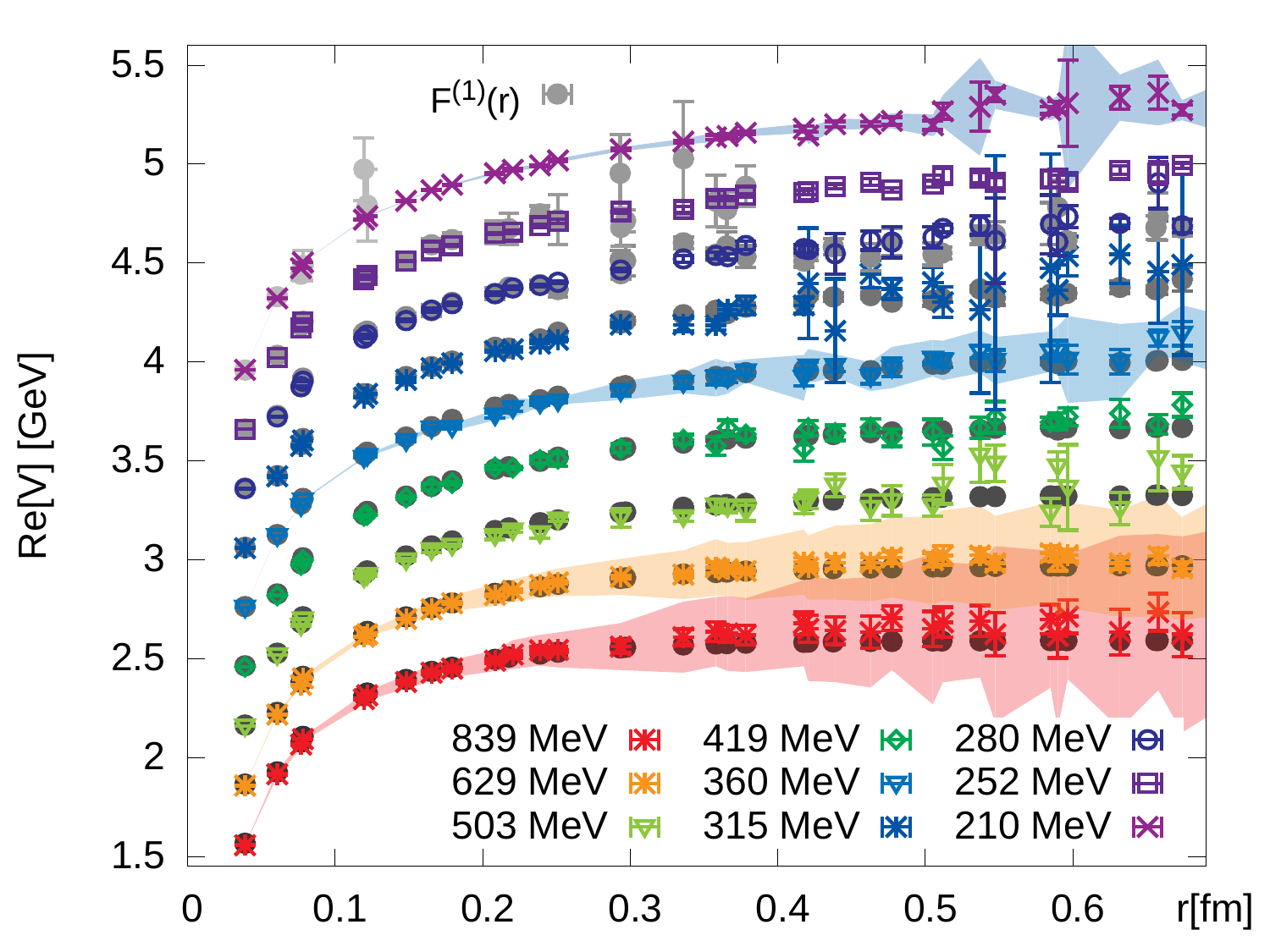}\hspace{1cm}
 \includegraphics[scale=0.4]{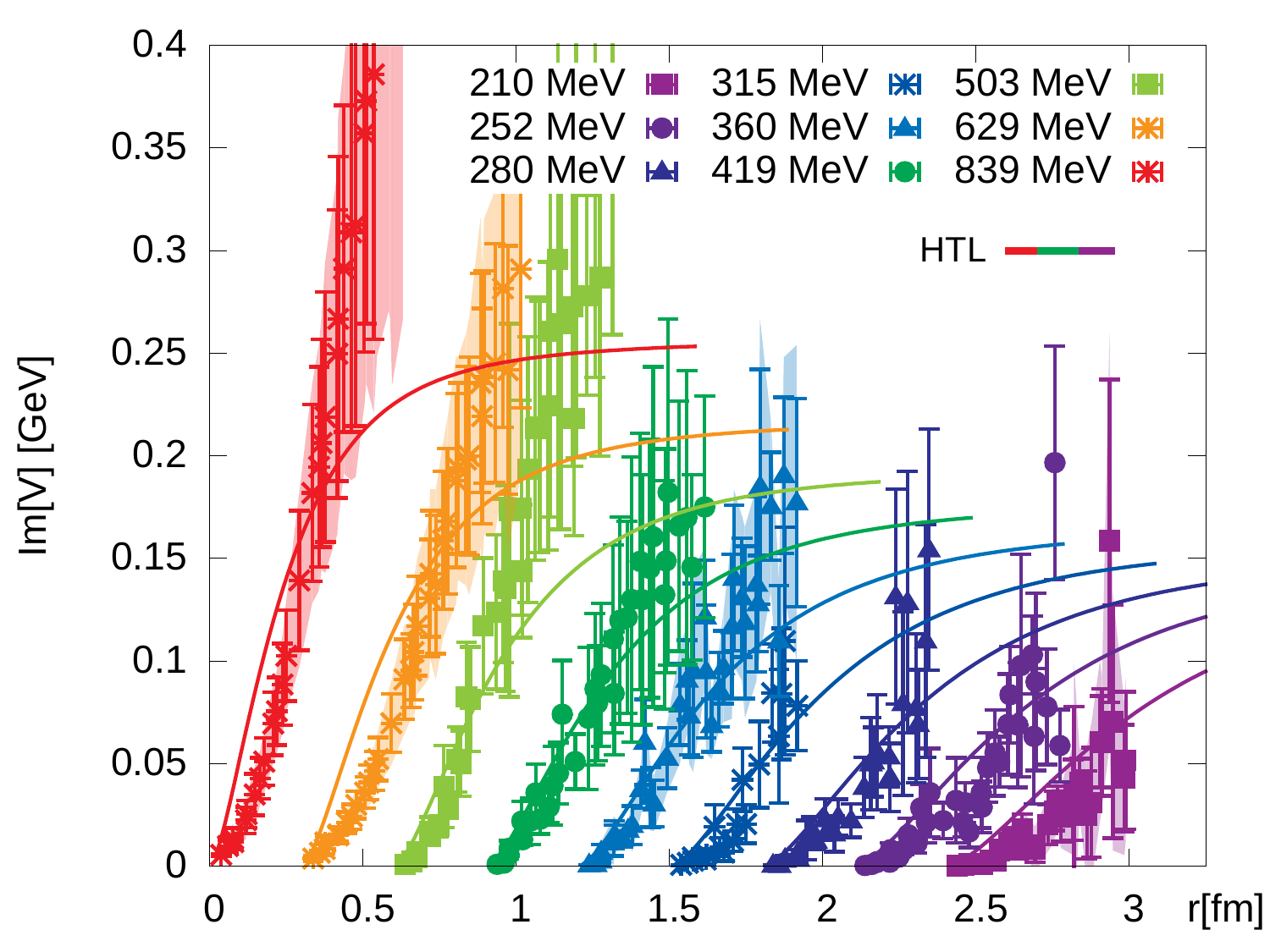}
 \caption{ (left) Real- and (left) imaginary part of the complex in-medium potential extracted from quenched QCD simulations.}\label{Fig:PotQuench}
\end{figure}

\section{The complex in-medium heavy quark potential from lattice QCD}
\label{sec4}

We now wish to address the question of how to determine the potential $V(r)$ acting between two static quarks at finite temperature non-perturbatively. One starts from an EFT description in terms of $Q\bar{Q}$ wave-functions called pNRQCD \cite{Brambilla:2004jw,Brambilla:2008cx}. The values of $V(r)$ are determined by the procedure of matching, in which a suitable QCD correlation function is identified, which carries the same physics content as a correlation functions in pNRQCD and both quantities are set equal at a certain scale. At $m_Q\to\infty$ the correlator of two color-singlet wavefunctions is identified with the real-time Wilson loop $W_\square(t,r)$. The potential arises from its late Minkowski-time limit 
\begin{align}
 W_\square(t,r)=\langle {\rm Tr}\Big({\rm exp}\big[ ig \int_\square dz^\mu A_\mu(z)]\Big)\rangle, \quad V(r)=\lim_{t\to\infty} \frac{i\partial_t W_\square(t,r)}{W_\square(t,r)}\label{Eq:Pot}
\end{align}
The limit reflects the fact that replacing an interaction mediated by gluons with an instantaneous potential is only valid at timescales much larger than those of the medium gluons. 

The challenge to a non-perturbatively evaluation lies in the real-time character of the definition, not directly amenable to a lattice QCD simulation. Here the concept of spectral function comes to our rescue \cite{Rothkopf:2009pk,Rothkopf:2011db} since the Wilson loop can be rewritten as a Fourier transform of a positive definite $\rho_\square(\omega,r)$. In fact the Euclidean Wilson loop is related to the same $\rho_\square$ via the Laplace transform, which can be inverted using the Bayesian method of sec.\ref{sec2}. Inserting the spectral decomposition in the potential definition \eqref{Eq:Pot} leads to
$V(r)=\lim_{t\to\infty}\int d\omega\, \omega e^{-i\omega t} \rho(\omega,r)/\int d\omega\, e^{-i\omega t} \rho(\omega,r). \label{Eq:PotSpec}$
The late time limit tells us that only the lowest lying spectral structure contributes to $V(r)$, which has been shown to take on the form of a skewed Lorentzian \cite{Burnier:2012az}. Its position and width correspond to the real- and imaginary part of the potential respectively. Hence we fit the lowest lying peak and read off ${\rm Re}[V]$ and ${\rm Im}[V]$.

In Fig.\ref{Fig:PotQuench} we show (colored points) the results \cite{Burnier:2014ssa,Burnier:2014yda} for (left) the real- and (right) imaginary part obtained from anisotropic $32^3\times N_\tau$ lattices with $\beta=7$ and bare anisotropy $\xi=3.5$ in the quenched approximation ($T_C=270$MeV). Changing $N_\tau=24-96$ we scan a range of $T=210-839$MeV. Note that instead of Wilson loops we here use Wilson line correlators in Coulomb gauge since they do not posses cusp divergences.

\begin{wrapfigure}{r}{0.4\textwidth}
  \begin{center}\vspace{-0.8cm}
   \includegraphics[scale=0.4]{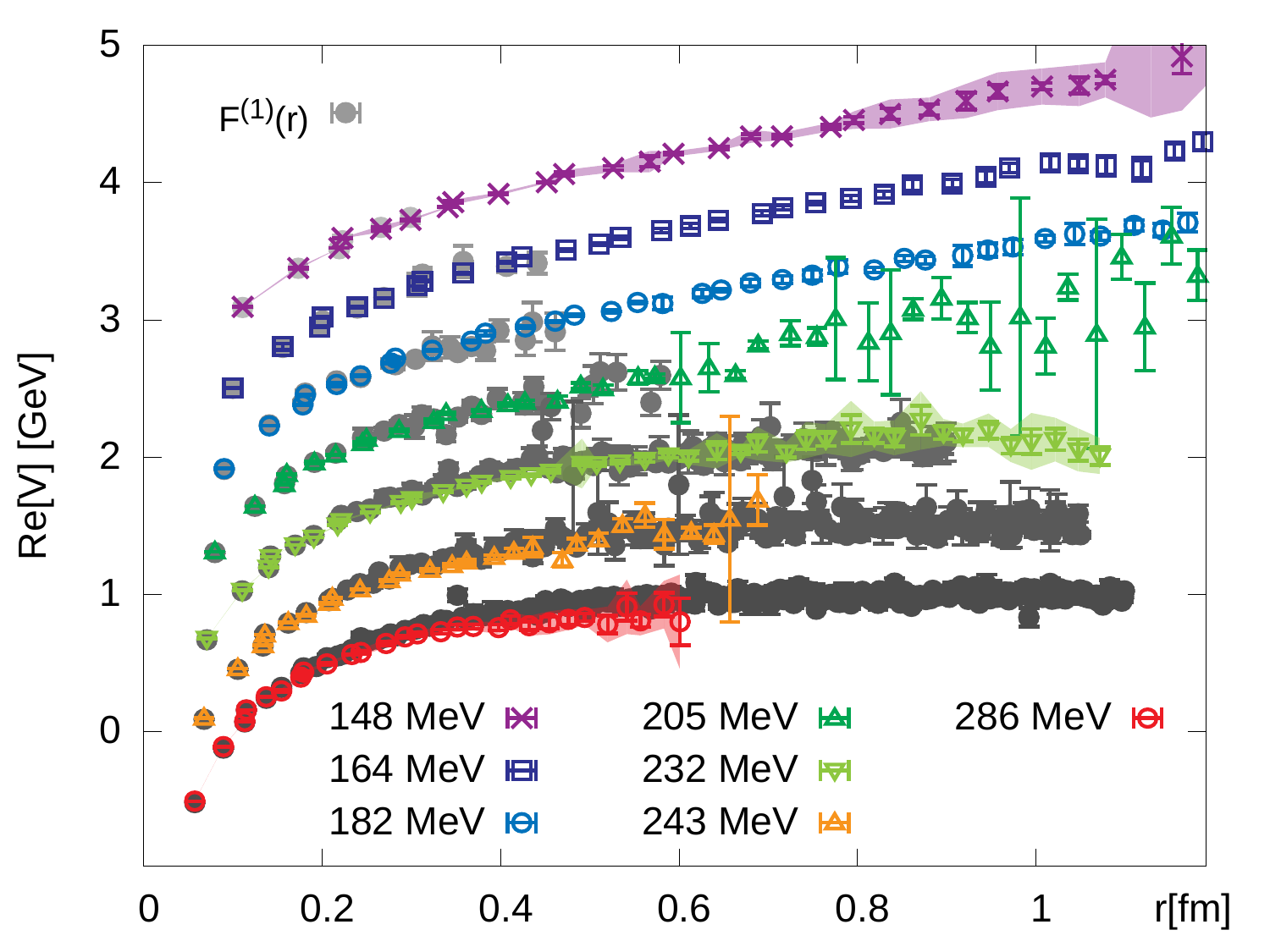}\vspace{-0.5cm}
  \end{center}
  \caption{ ${ \rm Re}[V]$ in full QCD with dynamical u,d and s quarks.}\label{Fig:PotFull}\vspace{-0.2cm}
\end{wrapfigure}

${\rm Re}[V]$ shows a smooth transition from the well known Cornell type behavior at low $T$ to a Debye screened form at $T>T_C$, as expected from perturbation theory. Interestingly the values of $F^{(1)}(r)$ (gray points) lie very close to the values of the real part. Hence this first principles calculation supports $F^{(1)}(r)$ as an approximation of ${\rm Re}[V]$. ${\rm Im}[V]$ is non-zero on the lattice at $T>T_C$ and hence cannot be neglected in an in-medium potential description. Its shooting off at larger $r$ is an artifact of the reconstruction due to a diminishing signal to noise ratio in the correlators.

We present in Fig.\ref{Fig:PotFull} the values of ${\rm Re}[V]$ in a true QGP with u,d, and s quarks. It is extracted on isotropic $48^3\times12$ lattices by the HotQCD collaboration \cite{Burnier:2014ssa,Burnier:2014yda}, spanning a temperature range of $T=148-286$MeV corresponding to a lattice parameter range of $\beta=6.8-7.48$. Based on the asqtad action with $m_\pi\simeq300$MeV the transition temperature here is $T=172.5$MeV. The behavior we find is qualitatively similar to the quenched case with a smooth transition from Cornell to Debye screened behavior, and again $F^{(1)}(r)$ appears to be a good approximation of ${\rm Re}[V]$. We refrain from showing the tentative values of ${\rm Im}[V]$, since spectral widths extracted from $N_\tau=12$ points are generally considered not reliable.

\section{Conclusion}

Heavy quarkonium presents a formidable testing ground for our capability to connect a first principles understanding of the strong interactions to phenomenologically relevant observables. The combination of non-relativistic EFT's with non-perturbative simulations of the QCD medium, already highly successful at $T\simeq0$, are nowadays providing valuable insight also at $T\simeq T_C$, reached in relativistic heavy ion-collisions. We reported on recent progress made in the determination of dynamical information from such simulations through a novel Bayesian spectral reconstruction approach and two studies based on its application were discussed. The first work extracted the in-medium spectra of bottomonium directly from lattice NRQCD and found survival of both S-wave and P-wave ground states up to $T=249$MeV. The second work used  spectral functions to extract the values of the complex in-medium heavy-quark potential. Besides a smooth transition from a confining to Debye screened behavior it confirmed the existence of an imaginary part in the potential in the quark-gluon-plasma. 

The author thanks Y. Burnier, O. Kaczmarek, S. Kim and P. Petreczky for the fruitful collaboration that underlies the lattice QCD spectra results presented in this proceeding.


\section*{References}
\begin{thebibliography}{9}

\bibitem{Brambilla:2010cs} 
  N.~Brambilla {\it et al.},
  Eur.\ Phys.\ J.\ C {\bf 71}, 1534 (2011)

\bibitem{Chatrchyan:2011pe}
  S.~Chatrchyan {\it et al.} [CMS],
  Phys.\ Rev.\ Lett.\  {\bf 107} (2011) 052302,Phys.Rev.Lett. {\bf 109} (2012) 222301 .


\bibitem{Abelev:2012rv} 
  B.~Abelev {\it et al.} [ALICE],
  Phys.\ Rev.\ Lett.\  {\bf 109}, 072301 (2012)
  
\bibitem{Andronic:2009sv} 
  A.~Andronic, F.~Beutler, P.~Braun-Munzinger, K.~Redlich and J.~Stachel,
  Phys.\ Lett.\ B {\bf 678}, 350 (2009)
  
\bibitem{Agashe:2014kda} 
  K.~A.~Olive {\it et al.} [Particle Data Group],
  Chin.\ Phys.\ C {\bf 38}, 090001 (2014).
  
\bibitem{Brambilla:2004jw} 
  N.~Brambilla, A.~Pineda, J.~Soto and A.~Vairo,
  Rev.\ Mod.\ Phys.\  {\bf 77}, 1423 (2005) 

\bibitem{Nadkarni:1986as} 
  S.~Nadkarni,
  Phys.\ Rev.\ D {\bf 34}, 3904 (1986);

\bibitem{Satz:2008zc} 
  H.~Satz,
  J.\ Phys.\ G {\bf 36}, 064011 (2009).
  
\bibitem{Wong:2004zr} 
  C.~Y.~Wong,
  Phys.\ Rev.\ C {\bf 72}, 034906 (2005);
  
\bibitem{Kaczmarek:2007pb} 
  O.~Kaczmarek,
  PoS CPOD {\bf 07}, 043 (2007);
  
\bibitem{Barchielli:1986zs} 
   A.~Barchielli, E.~Montaldi and G.~M.~Prosperi,
   Nucl.\ Phys.\ B {\bf 296}, 625 (1988),
  
\bibitem{Brambilla:2008cx} 
  N.~Brambilla, J.~Ghiglieri, A.~Vairo, P.~Petreczky,
  Phys.\ Rev.\ D {\bf 78}, 014017 (2008) 
      

\bibitem{Laine:2007qy} 
  M.~Laine, O.~Philipsen and M.~Tassler,
  JHEP {\bf 0709}, 066 (2007) 
  
\bibitem{Beraudo:2007ky} 
  A.~Beraudo, J.~-P.~Blaizot, C.~Ratti,
  Nucl.\ Phys.\ A {\bf 806}, 312 (2008) 
  

\bibitem{Burnier:2014ssa} 
  Y.~Burnier, O.~Kaczmarek and A.~Rothkopf,
  Phys.\ Rev.\ Lett.\  {\bf 114}, no. 8, 082001 (2015)
  
\bibitem{Burnier:2014yda} 
  Y.~Burnier, O.~Kaczmarek and A.~Rothkopf,
  PoS LATTICE {\bf 2014}, 220 (2015)

\bibitem{Thacker:1990bm} 
  B.~A.~Thacker and G.~P.~Lepage,
  Phys.\ Rev.\ D {\bf 43}, 196 (1991).
  
\bibitem{Dowdall:2011wh} 
  R.~J.~Dowdall {\it et al.} [HPQCD],
  Phys.\ Rev.\ D {\bf 85}, 054509 (2012)
  

\bibitem{Mizuk:2012pb} 
  R.~Mizuk {\it et al.} [Belle],
  Phys.\ Rev.\ Lett.\  {\bf 109}, 232002 (2012)


  
\bibitem{Aoki:2006we} 
  Y.~Aoki, G.~Endrodi, Z.~Fodor, S.~D.~Katz and K.~K.~Szabo,
  Nature {\bf 443}, 675 (2006)
 
\bibitem{Bazavov:2011nk} 
  A.~Bazavov {\it et al.},
  Phys.\ Rev.\ D {\bf 85}, 054503 (2012)
  
\bibitem{Bazavov:2014pvz} 
  A.~Bazavov {\it et al.} [HotQCD],
  Phys.\ Rev.\ D {\bf 90}, no. 9, 094503 (2014)

\bibitem{Borsanyi:2013bia} 
  S.~Borsanyi, Z.~Fodor, C.~Hoelbling, S.~D.~Katz, S.~Krieg and K.~K.~Szabo,
  Phys.\ Lett.\ B {\bf 730}, 99 (2014)
 
  
\bibitem{Burger:2014xga} 
  F.~Burger {\it et al.} [tmfT],
  Phys.\ Rev.\ D {\bf 91}, no. 7, 074504 (2015)
  
\bibitem{Asakawa:2000tr}
  M.~Asakawa, T.~Hatsuda and Y.~Nakahara,
  Prog.\ Part.\ Nucl.\ Phys.\  {\bf 46} (2001) 459 
  
\bibitem{Jakovac:2006sf} 
  A.~Jakovac, P.~Petreczky, K.~Petrov and A.~Velytsky,
  Phys.\ Rev.\ D {\bf 75}, 014506 (2007)
 
\bibitem{Rothkopf:2011ef} 
  A.~Rothkopf,
  J.\ Comput.\ Phys.\  {\bf 238}, 106 (2013)
 
\bibitem{Burnier:2013nla} 
  Y.~Burnier and A.~Rothkopf,
  Phys.\ Rev.\ Lett.\  {\bf 111}, 182003 (2013),PoS LATTICE {\bf 2013}, 490 (2014)
  
\bibitem{Davies:1991py} 
  C.~T.~H.~Davies and B.~A.~Thacker,
  Phys.\ Rev.\ D {\bf 45}, 915 (1992).
  
\bibitem{Aarts:2011sm} 
  G.~Aarts, {\it et al.} [FASTSUM] 
  JHEP {\bf 1111}, 103 (2011),JHEP {\bf 1312}, 064 (2013) ,JHEP {\bf 1407}, 097 (2014)
  
\bibitem{Burnier:2007qm} 
  Y.~Burnier, M.~Laine and M.~Vepsalainen,
  JHEP {\bf 0801}, 043 (2008)
  

  

\bibitem{Kim:2014iga} 
  S.~Kim, P.~Petreczky and A.~Rothkopf,
  Phys.\ Rev.\ D {\bf 91}, 054511 (2015),PoS LATTICE {\bf 2014}, 208 (2014)
  
  
\bibitem{Rothkopf:2009pk} 
  A.~Rothkopf, T.~Hatsuda and S.~Sasaki,
  PoS LAT {\bf 2009}, 162 (2009)
  
\bibitem{Rothkopf:2011db} 
  A.~Rothkopf, T.~Hatsuda and S.~Sasaki,
  Phys.\ Rev.\ Lett.\  {\bf 108}, 162001 (2012)
  
\bibitem{Burnier:2012az} 
  Y.~Burnier and A.~Rothkopf,
  Phys.\ Rev.\ D {\bf 86}, 051503 (2012)
   
  
\end{thebibliography}
\end{document}